\documentclass{article}
\usepackage{subcaption}
\usepackage{spconf}
\usepackage{amsmath}
\usepackage{amssymb}
\usepackage{graphicx}
\usepackage{tikz}
\usetikzlibrary{patterns, arrows, automata}
\usepackage{pgfplots}
\usepgfplotslibrary{fillbetween}
\pgfplotsset{compat=1.5}
\pgfplotsset{major grid style={dashed}}
\usepackage[sorting=nty, style=ieee, citestyle=ieee]{biblatex}
\usepackage{algorithm}
\usepackage{algpseudocode}
\usepackage{algorithmicx}

\definecolor{SoftGreen}{HTML}{2FBB99}
\definecolor{SoftBlue}{HTML}{2F97BA}
\definecolor{SoftRed}{HTML}{F75D59}

\title{AN ADAPTIVE PRUNING ALGORITHM FOR SPOOFING LOCALISATION BASED ON TROPICAL GEOMETRY}
\name{Emmanouil Theodosis and Petros Maragos}
\address{School of ECE, National Technical University of Athens, Athens, Greece\\
manostheodosis@gmail.com, maragos@cs.ntua.gr}
\begin{document}
\ninept
\maketitle
\begin{abstract}
The problem of spoofing attacks is increasingly relevant as digital systems are becoming more ubiquitous. Thus the detection of such attacks and the localisation of attackers have been objects of recent study. After an attack has been detected, various algorithms have been proposed in order to localise the attacker. In this work we propose a new adaptive pruning algorithm inspired by the tropical and geometrical analysis of the traditional Viterbi pruning algorithm to solve the localisation problem. In particular, the proposed algorithm tries to localise the attacker by adapting the leniency parameter based on estimates about the state of the solution space. These estimates stem from the enclosed volume and the entropy of the solution space, as they were introduced in our previous works.
\end{abstract}
\section{Introduction}
\label{sec:intro}
Spoofing attacks have been the object of study of various researchers as computer systems became more prevalent. In these attacks, a malicious individual aims to gain access to a system's resources by masking their true identity and intentions, in order to either inflict damage to the system or access unauthorised content. Such attacks can be remote attacks to a computer networks (\cite{CTM07}, \cite{LiTr06a}, \cite{LiTr06b}, \cite{PDB13}), but can also have a physical aspect (\cite{KMD17}, \cite{SKP18}). The overaching goal of the field is to design systems that are able to efficiently detect and localise such attacks, without infringing significant overheads to the system, and without hindering the overall experience of the users. To the effect of localisation, the employment of various algorithms has been proposed, including clustering algorithms.

Pruning algorithms are ubiquitous in computer science and are applied in a vast array of problems, however their application to the localisation problem has been limited. Such algorithms are used in order emphasize the speed of computation over the optimality of the solution. This is essential in applications where real time computation requirements are indispensable (as is the detection of an attack). Several authors have proposed adaptive algorithms with varying motivations: some algorithms try to minimise the power consumption of the pruning procedure (\cite{Tes+05}, \cite{HeCh02}, \cite{HeCh04}), while others aim to employ ideas from theoretical computer science to improve the accuracy of the algorithm (\cite{BiLi10}, \cite{LiHa10}). Some authors even use techniques from control theory to adaptively alter the pruning parameter (\cite{ZhDu04}).

Adaptive algorithms are a prime candidate for analysis using tropical geometry (\cite{MaSt15}), which has been increasing in popularity. Many authors (\cite{ThMa18}, \cite{ChMa17}) resort to using tropical geometry for its appealing properties; namely the piecewise linearity of the solution space, and the intuitive reasoning regarding that space. Tropical geometry allows for a layer of abstraction; instead of reasoning about the algorithm itself, we can reason about the solution space it produces, which often can lead to deductions about the possible solutions. This can be exteremely useful in modern day, since the sheer size and dimensionality of the input data can make explicit remarks about the function of algorithms near impossible.

Despite tropical geometry's appeal, previous approaches of adaptive pruning algorithms have not taken advantage of these properties. Authors have proposed (numerical) optimisations which can reduce the energy consumption of convolutional code decoding, and others have analysed the structure of specific models in order to make deductions and predictions about the pruning parameter. In contrast, in this paper we try to adapt the pruning parameter based solely on the shape and state of the solution space, without assuming any specific structure of the applied model.

Reference \cite{CTM07} proposes the use of the K-means algorithm for the detection and localisation of spoofing attacks. Extended work has been done on the detection of spoofing attacks, both in communication networks (\cite{LiTr06a}, \cite{LiTr06b}), but also in speech recognition systems (\cite{KMD17}, \cite{SKP18}). In \cite{PDB13} the authors try to offer a mathematical framework for cyber attacks from a system-theoretic perspective. References \cite{Tes+05}, \cite{HeCh02}, and \cite{HeCh04} tackle pruning from a telecommunications perspective, aiming to minimize the energy consumption during decoding in receivers. The authors of \cite{ZhDu04} aim to utilise metrics, derived from an assumed system structure, to predict the evolution of the leniency parameter. Reference \cite{BiLi10} tries to exploit the inherent nature of speech recognition in order to speed up pruning. Finally, in \cite{LiHa10} the authors try to efficiently understand the structure of the solution space by computing cliques in order to, subsequently, improve pruning. However, that approach is supervised; each specific application domain has to be analysed and evaluated as to whether cliques can be computed, and thus the approach is not generilisable.

In this work we propose an adaptive variation of the Viterbi pruning in order to solve the localisation problem by exploiting the geometrical structure of the solution space. In particular, tropical polytopes can be defined during each step of the Viterbi algorithm. We utilise two metrics (defined in our previous work \cite{ThMa18}) deriving from the tropical polytopes of the Viterbi pruning in order to design the new adaptive algorithm. The proposed algorithm computes the metrics' values at each time frame and then compares them with a previous history in order to decide whether or not to adapt the current value of the pruning parameter. In the case that pruning is indeed warranted, the algorithm tries to adapt the parameter to the effect of maintaining the volume enclosed in the solution space.

In Section \ref{sec:back} we introduce the background upon which this work is based. Section \ref{sec:algo} presents the proposed adaptive algorithm and briefly analyses its function. Finally, in Section \ref{sec:res} we apply the proposed algorithm to a simulated attack on a network.

\section{Background}
\label{sec:back}
\subsection{Tropical Algebra and Geometry}
Tropical algebra (\cite{GoMi08}, \cite{Butk10}, \cite{Cuni79}) is an algebraic body similar to linear algebra, where the pair of main operations is $(\wedge, +)$. It operates on the extended real multidimentional space $\mathbb{R}_{\min}^n$ ($\mathbb{R}_{\min} = \mathbb{R} \cup \{+\infty\}$). The min-plus matrix multiplication is denoted $\boxplus$, and its result between two matrices $\mathbf{A}, \mathbf{B} \in \mathbb{R}_{\min}^n$ is given by:
\begin{equation}
    \label{eq:prod}
    \left( \mathbf{A} \boxplus \mathbf{B} \right)_{ij} = \bigwedge_{k=1}^{n}A_{ik} + B_{kj}
\end{equation}
where $\wedge$ denotes the minimum (see \cite{Mara17} for details).

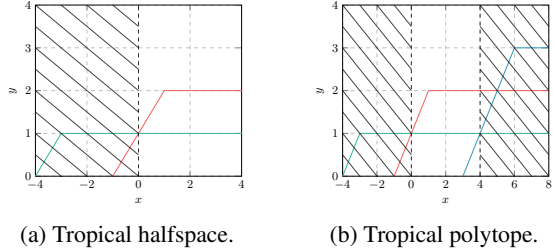
\begin{figure}[h]
    \begin{subfigure}[b]{0.225\textwidth}
        \begin{center}
        \begin{tikzpicture}[scale=0.5]
            \begin{axis}[
                    xlabel={$x$},
                    ylabel={$y$},
                    ymin=0, ymax=4,
                    xmin=-4, xmax=4,
                    grid=major,
                    scale=0.8,
                ]
                \draw[SoftGreen] (axis cs: -4, 0) -- (axis cs: -3, 1) -- (axis cs: 5, 1);

                \draw[SoftRed] (axis cs: -2, -1) -- (axis cs: 1, 2) -- (axis cs: 5, 2);

                \draw[dashed, opacity=0.75] (axis cs: 0, 0) -- (axis cs: 0, 4);

                \draw[opacity=0.75] (axis cs: -4, 0.5) -- (axis cs: 0, -1.5);
                \draw[opacity=0.75] (axis cs: -4, 1) -- (axis cs: 0, -1);
                \draw[opacity=0.75] (axis cs: -4, 1.5) -- (axis cs: 0, -0.5);
                \draw[opacity=0.75] (axis cs: -4, 2) -- (axis cs: 0, 0);
                \draw[opacity=0.75] (axis cs: -4, 2.5) -- (axis cs: 0, 0.5);
                \draw[opacity=0.75] (axis cs: -4, 3) -- (axis cs: 0, 1);
                \draw[opacity=0.75] (axis cs: -4, 3.5) -- (axis cs: 0, 1.5);
                \draw[opacity=0.75] (axis cs: -4, 4) -- (axis cs: 0, 2);
                \draw[opacity=0.75] (axis cs: -4, 4.5) -- (axis cs: 0, 2.5);
                \draw[opacity=0.75] (axis cs: -4, 5) -- (axis cs: 0, 3);
                \draw[opacity=0.75] (axis cs: -4, 5.5) -- (axis cs: 0, 3.5);
            \end{axis}
        \end{tikzpicture}
        \caption{Tropical halfspace.}
    \end{center}
    \end{subfigure}
    \begin{subfigure}[b]{0.225\textwidth}
        \begin{center}
        \begin{tikzpicture}[scale=0.5]
            \begin{axis}[
                    xlabel={$x$},
                    ylabel={$y$},
                    ymin=0, ymax=4,
                    xmin=-4, xmax=8,
                    grid=major,
                    scale=0.8,
                ]
                \draw[SoftGreen] (axis cs: -4, 0) -- (axis cs: -3, 1) -- (axis cs: 8, 1);
                \draw[SoftRed] (axis cs: -2, -1) -- (axis cs: 1, 2) -- (axis cs: 8, 2);

                \draw[dashed, opacity=0.75] (axis cs: 0, 0) -- (axis cs: 0, 6);

                \draw[opacity=0.75] (axis cs: -4, 0.5) -- (axis cs: 0, -1.5);
                \draw[opacity=0.75] (axis cs: -4, 1) -- (axis cs: 0, -1);
                \draw[opacity=0.75] (axis cs: -4, 1.5) -- (axis cs: 0, -0.5);
                \draw[opacity=0.75] (axis cs: -4, 2) -- (axis cs: 0, 0);
                \draw[opacity=0.75] (axis cs: -4, 2.5) -- (axis cs: 0, 0.5);
                \draw[opacity=0.75] (axis cs: -4, 3) -- (axis cs: 0, 1);
                \draw[opacity=0.75] (axis cs: -4, 3.5) -- (axis cs: 0, 1.5);
                \draw[opacity=0.75] (axis cs: -4, 4) -- (axis cs: 0, 2);
                \draw[opacity=0.75] (axis cs: -4, 4.5) -- (axis cs: 0, 2.5);
                \draw[opacity=0.75] (axis cs: -4, 5) -- (axis cs: 0, 3);
                \draw[opacity=0.75] (axis cs: -4, 5.5) -- (axis cs: 0, 3.5);
                \draw[opacity=0.75] (axis cs: -4, 6) -- (axis cs: 0, 4);
                \draw[opacity=0.75] (axis cs: -4, 6.5) -- (axis cs: 0, 4.5);
                \draw[opacity=0.75] (axis cs: -4, 7) -- (axis cs: 0, 5);
                \draw[opacity=0.75] (axis cs: -4, 7.5) -- (axis cs: 0, 5.5);
                \draw[opacity=0.75] (axis cs: -4, 8) -- (axis cs: 0, 6);

                \draw[SoftBlue] (axis cs: 3, 0) -- (axis cs: 6, 3) -- (axis cs: 8, 3);

                \draw[dashed, opacity=0.75] (axis cs: 4, 0) -- (axis cs: 4, 6);

                \draw[opacity=0.75] (axis cs: 4, 0.5) -- (axis cs: 8, -1.5);
                \draw[opacity=0.75] (axis cs: 4, 1) -- (axis cs: 8, -1);
                \draw[opacity=0.75] (axis cs: 4, 1.5) -- (axis cs: 8, -0.5);
                \draw[opacity=0.75] (axis cs: 4, 2) -- (axis cs: 8, 0);
                \draw[opacity=0.75] (axis cs: 4, 2.5) -- (axis cs: 8, 0.5);
                \draw[opacity=0.75] (axis cs: 4, 3) -- (axis cs: 8, 1);
                \draw[opacity=0.75] (axis cs: 4, 3.5) -- (axis cs: 8, 1.5);
                \draw[opacity=0.75] (axis cs: 4, 4) -- (axis cs: 8, 2);
                \draw[opacity=0.75] (axis cs: 4, 4.5) -- (axis cs: 8, 2.5);
                \draw[opacity=0.75] (axis cs: 4, 5) -- (axis cs: 8, 3);
                \draw[opacity=0.75] (axis cs: 4, 5.5) -- (axis cs: 8, 3.5);
                \draw[opacity=0.75] (axis cs: 4, 6) -- (axis cs: 8, 4);
                \draw[opacity=0.75] (axis cs: 4, 6.5) -- (axis cs: 8, 4.5);
                \draw[opacity=0.75] (axis cs: 4, 7) -- (axis cs: 8, 5);
                \draw[opacity=0.75] (axis cs: 4, 7.5) -- (axis cs: 8, 5.5);
                \draw[opacity=0.75] (axis cs: 4, 8) -- (axis cs: 8, 6);
            \end{axis}
        \end{tikzpicture}
        \caption{Tropical polytope.}
    \end{center}
    \end{subfigure}
    \caption{Tropical halfspaces and tropical polytopes are the result of the tropicalisation of their Euclidean counterparts.}
    \label{fig:polytopes}
\end{figure}

Tropical geometry (\cite{MaSt15}) studies the objects of Euclidean geometry under the tropical prism. Similar to its Euclidean counterpart, a tropical polytope will be a closed intersection of a finite number of tropical halfspaces. Figure \ref{fig:polytopes} offers visual examples of tropical halfspaces and polytopes.

\subsection{Tropical Viterbi}
The Viterbi algorithm can be written in tropical algebra, as we proposed in \cite{ThMa18}, in the following closed form:
\begin{equation}
    \label{eq:viterbi}
    \mathbf{x}(t) = \mathbf{P}(\sigma_t) \boxplus \mathbf{A}^T \boxplus \mathbf{x}(t - 1)
\end{equation}
where $\mathbf{x}(t)$ is the state vector, $\mathbf{A}$ is the matrix of the transition weights, and $\mathbf{P}(\sigma_t)$ is a diagonal matrix containing the observation weights for the input symbol $\sigma_t$ at each state.

In \cite{ThMa18} we analysed the pruning variant of the Viterbi algorithm in tropical algebra and comment on its geometry. Therein, a vector of variables $\mathbf{z}$ is considered, and then it is bounded by the Viterbi update law of \eqref{eq:viterbi} and the pruning vector $\pmb{\eta} = \theta + \frac{1}{2} \left( \mathbf{x}(t)^T \boxplus \mathbf{x}(t) \right) + \mathbf{0}$, where $\theta$ is the leniency parameter. This defines a tropical polytope on the variable vector $\mathbf{z}$, which encloses all the possible assignments of the variables that satisfy the constraints (which, essentially, is the solution space for the pruning procedure). Then, at every interval of the algorithm, two metrics are calculated based on that polytope:
\begin{itemize}
    \item a metric based on the normalized volume inside the polytope:
    \begin{equation}
        \label{eq:nu}
        \nu = -\frac{1}{\mathrm{supp}(\mathbf{z})} \sum_{i \in \mathrm{supp}(\mathbf{z})} \frac{\log r_i}{\log \left( \max \mathbf{r}\right)}
    \end{equation}
    \item a metric based on the entropy of the polytope:
    \begin{equation}
        \label{eq:varepsilon}
        \varepsilon = -\frac{1}{\mathrm{supp}(\mathbf{z})} \sum_{i \in \mathrm{supp}(\mathbf{z})}{-z_i(t) \cdot e^{-z_i(t)}}
    \end{equation}
\end{itemize}
where $r_i = \eta - z_i$. Essentially, $r_i$ is the degree to which each dimension satisfies the Viterbi constraints.

\subsection{Poisson distributions}
It is very common for network requests in telecommunications applications to be modeled as Poisson distributions (\cite{BeTs08}, \cite{Pitm99}). Poisson distributions are controlled by the parameter $\lambda$, which can be interpreted as the mean amount of requests in a time frame. Alternatively, in queueing systems the $\lambda$ parameter can be perceived as the average time units that a user will have to wait until he is serviced. Formally, the probability of witnessing exactly $\kappa$ requests in the time frame (or waiting for $\kappa$ time units) is given by Equation \eqref{eq:poisson}:

\begin{equation}
    \label{eq:poisson}
    P[X = \kappa] = e^{-\lambda}\frac{\lambda^\kappa}{\kappa!}
\end{equation}
where $X$ is the random variable modeling the number of requests in the time frame. Some important characteristics of the distribution:
\begin{itemize}
    \item The distribution is discrete-valued. This makes it ideal to model number of requests in a network.
    \item The distribution's mean $\mu$ is equal to the parameter $\lambda$.
    \item The distribution's variance $\sigma^2$ is equal to the parameter $\lambda$.
    \item The distribution is \emph{memoryless}. In essence, this means that if a user has already made $\tau$ requests, then the probability of making a total of $t + \tau$ requests is equal to the probability of making $t$ requests. Formally:
    \begin{equation}
        \label{eq:memory}
        P[X > t + \tau | X \geq \tau] = P[X > t]
    \end{equation}
\end{itemize}

\section{Algorithm}
\label{sec:algo}
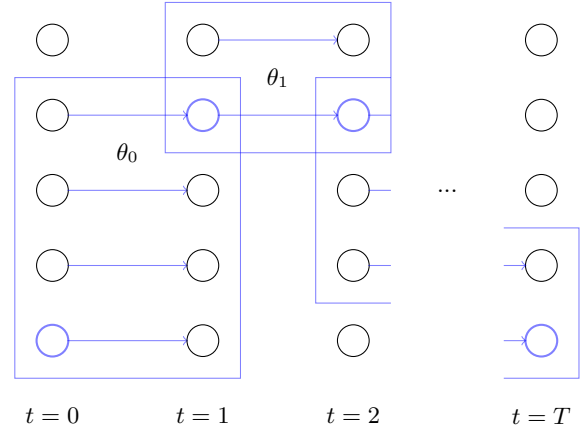
\begin{figure}[t]
    \begin{center}
        \begin{tikzpicture}
            \draw (-2.5, 5) circle (6pt);
            \draw (-2.5, 4) circle (6pt);
            \draw (-2.5, 3) circle (6pt);
            \draw (-2.5, 2) circle (6pt);
            \draw[thick, blue, opacity=0.5] (-2.5, 1) circle (6pt);
            \node (0) at (-2.5, 0) {$t = 0$};

            \draw[->, blue, opacity=0.5] (-2.3, 4) -- (-0.7, 4);
            \draw[->, blue, opacity=0.5] (-2.3, 3) -- (-0.7, 3);
            \draw[->, blue, opacity=0.5] (-2.3, 2) -- (-0.7, 2);
            \draw[->, blue, opacity=0.5] (-2.3, 1) -- (-0.7, 1);
            \node (5) at (-1.5, 3.5) {$\theta_0$};

            \draw[blue, opacity=0.5] (-3, 4.5) -- (0, 4.5) -- (0, 0.5) -- (-3, 0.5) -- cycle;

            \draw (-0.5, 5) circle (6pt);
            \draw[thick, blue, opacity=0.5] (-0.5, 4) circle (6pt);
            \draw (-0.5, 3) circle (6pt);
            \draw (-0.5, 2) circle (6pt);
            \draw (-0.5, 1) circle (6pt);
            \node (1) at (-0.5, 0) {$t = 1$};

            \draw[->, blue, opacity=0.5] (-0.3, 5) -- (1.3, 5);
            \draw[->, blue, opacity=0.5] (-0.3, 4) -- (1.3, 4);
            \node (6) at (0.5, 4.5) {$\theta_1$};

            \draw[blue, opacity=0.5] (-1, 5.5) -- (2, 5.5) -- (2, 3.5) -- (-1, 3.5) -- cycle;

            \draw (1.5, 5) circle (6pt);
            \draw[thick, blue, opacity=0.5] (1.5, 4) circle (6pt);
            \draw (1.5, 3) circle (6pt);
            \draw (1.5, 2) circle (6pt);
            \draw (1.5, 1) circle (6pt);
            \node (2) at (1.5, 0) {$t = 2$};

            \draw[blue, opacity=0.5] (1.7, 4) -- (2, 4);
            \draw[blue, opacity=0.5] (1.7, 3) -- (2, 3);
            \draw[blue, opacity=0.5] (1.7, 2) -- (2, 2);

            \draw[blue, opacity=0.5] (2, 1.5) -- (1, 1.5) -- (1, 4.5) -- (2, 4.5);

            \node (3) at (2.75, 3) {$...$};

            \draw (4, 5) circle (6pt);
            \draw (4, 4) circle (6pt);
            \draw (4, 3) circle (6pt);
            \draw (4, 2) circle (6pt);
            \draw[thick, blue, opacity=0.5] (4, 1) circle (6pt);
            \node (4) at (4, 0) {$t = T$};

            \draw[->, blue, opacity=0.5] (3.5, 2) -- (3.8, 2);
            \draw[->, blue, opacity=0.5] (3.5, 1) -- (3.8, 1);

            \draw[blue, opacity=0.5] (3.5, 2.5) -- (4.5, 2.5) -- (4.5, 0.5) -- (3.5, 0.5);
        \end{tikzpicture}
    \end{center}
    \caption{The proposed algorithm calculates a new value $\theta_i$ for the leniency parameter $\theta$ of each time frame. Based on the entropy and the volume of the solution space at that time frame, the value $\theta_i$ is adapted to allow for the survival of more, or fewer, paths.}
    \label{fig:algo}
\end{figure}

We propose a novel adaptive pruning algorithm that dynamically adapts the pruning parameter by consulting the metrics $\nu$ and $\varepsilon$ (\cite{ThMa18}). The algorithm computes the metrics and then decides, based on a history of previous values, if the current interval warrants an adaptation of the pruning parameter $\theta$. If that proves to be the case, the $\theta$ parameter is increased or decreased accordingly, in order to maintain the volume enclosed in the polytope. In essence, the algorithm computes a new value $\theta_i$ for each time frame, based on the state of the solution space (Figure \ref{fig:algo}). The main parameters of the algorithm are:

i. the parameter $\alpha$, which is the percentage threshold for $\varepsilon$. If the current value of $\varepsilon$ differs from the running history by a percentage more than $\alpha$, then the algorithm will adapt the parameter $\theta$.

ii. the parameter $\beta$, which is the percentage change for $\theta$. If the algorithm proceeds to the adaptation of $\theta$, the current value of $\nu$ is compared with the running history. If it is larger, then this means that the current value of $\theta$ allows for the inclusion of more paths than before, and thus the algorithms proceeds to decrease the value of $\theta$ by a percentage of $\beta$. Similarly, if $\nu$ is smaller than the running history, the algorithm increases $\theta$ by a percentage of $\beta$, in order to allow for the survival of more paths.

iii. the parameter $\tau$, which is the length of the running history. The algorithm first collects $\tau$ samples for the running history, and then compares the current interval's metrics with the average of the most recent $\tau$ entries in order to decide if an adaptation is warranted, and also whether to increase or decrease the parameter $\theta$.

iv. the parameter $\theta_0$, which is the initial value of the pruning parameter $\theta$. $\theta_0$ is also used as the pruning parameter for the first $\tau$ intervals, in order to calculate the history of the metrics $\varepsilon$ and $\nu$.

Algorithm \ref{algo:adapt} presents the proposed algorithm. Besides the parameters mentioned above, the algorithm also accepts other inputs required for the simulation and the Viterbi computation. In particular, a simulation runtime $T$ is required, as well as the number $n$ of states and the initial, transition, and observation costs $I, A,$ and $\Lambda$.

\begin{algorithm}[t]
    \caption{AdaptivePruning($T$, $n$, $I$, $A$, $\Lambda$, $\alpha$, $\beta$, $\tau$, $\theta_0$)}
    \label{algo:adapt}
    \begin{algorithmic}[1]
        \State $t \gets 0$
        \State $E, N, L \gets \emptyset$
        \State $\theta \gets \theta_0$
        \While{$t < T$}
            \State $\varepsilon, \nu \gets \textrm{polytope}(n, I, A, \Lambda, \theta)$
            \If{$t \geq \tau$}
                \If{$\frac{\lvert \varepsilon - \textrm{mean}(E) \rvert}{\textrm{mean}(E)} \geq \alpha$}
                    \If{$\nu \leq \textrm{mean}(N)$}
                        \State $\theta \gets (1 + \beta) \times \theta$
                    \Else
                        \State $\theta \gets (1 - \beta) \times \theta$
                    \EndIf
                \EndIf
            \EndIf
            \State $\textrm{update}(E, N, L)$
        \EndWhile
        \State $\textrm{seq} \gets \textrm{backtrack}(L)$
        \State \Return $\textrm{seq}, L$
    \end{algorithmic}
\end{algorithm}

In essence, the algorithm operates as follows. First, the Viterbi computation is performed using the previous value for the parameter $\theta$. Then, the geometrical metrics $\varepsilon$ and $\nu$ are calculated from the polytope of the Viterbi computation. Then, the algorithm compares the current value of the metric $\varepsilon$ with the running history. This is done because $\varepsilon$, essentially, calculates the entropy of the solution space. In information theory (\cite{ShWe98}, \cite{CoTh06}, \cite{Ston15}), entropy is a measure of surprise. In particular entropy is used to communicate if a new sample conveys a significant amount of information. For example, if a sample abides by the expectations of the current parameterisation of the distribution, then the entropy will be low, indicating that the sample provides no new information regarding the understanding of the distribution (and thus we are not surprised to observe this sample when we are sampling our distribution). Conversely, if a sample is atypical for the current parameterisation, then the entropy will be high, indicating that the sample provides new information regarding our understanding of the distribution, suggesting that the parameterisation might be wrong. The algorithm tries to leverage this measure of surprise, by comparing the level of current entropy with a running history. If there is a significant difference, this indicates a change in the solution space; the previous parameterisation of $\theta$ will no longer have similar effects. Thus, in such cases, the algorithm decides to update the value of $\theta$ in order to curb the levels of excitement.

\begin{figure}[t]
    \begin{center}
        \begin{tikzpicture}
            \draw[thick, opacity=0.25] (-1.4, 3) circle (10pt);
            \draw[thick, opacity=0.25] (-1.4, 2) circle (10pt);
            \draw[thick, opacity=0.25] (-1.4, 1) circle (10pt);
            \node[opacity=0.5] (0) at (-1.35, 0.25) {$\theta_{i-1}$};

            \draw[thick] (0.3, 3) circle (10pt);
            \draw[thick] (0.3, 2) circle (10pt);
            \draw[thick] (0.3, 1) circle (10pt);
            \node (0) at (0.3, 0.25) {$\theta_{i}$};

            \draw[thick, opacity=0.25] (2, 3) circle (10pt);
            \draw[thick, opacity=0.25] (2, 2) circle (10pt);
            \draw[thick, opacity=0.25] (2, 1) circle (10pt);
            \node[text=blue, opacity=0.8] (0) at (2, 3.6) {$\theta_{i+1}$};

            \draw[->, opacity=0.25] (-1.05, 3) -- (-0.05, 3);
            \draw[->, opacity=0.25] (-1.05, 3) -- (-0.05, 2);
            \draw[->, opacity=0.25] (-1.05, 3) -- (-0.05, 1);
            \draw[->, opacity=0.25] (-2, 3) -- (-1.75, 3);
            \draw[->, opacity=0.25] (-2, 2.8) -- (-1.75, 3);
            \draw[->, opacity=0.25] (-2, 2.6) -- (-1.75, 3);

            \draw[->, opacity=0.25] (-1.05, 2) -- (-0.05, 3);
            \draw[->, opacity=0.25] (-1.05, 2) -- (-0.05, 2);
            \draw[->, opacity=0.25] (-1.05, 2) -- (-0.05, 1);
            \draw[->, opacity=0.25] (-2, 2.2) -- (-1.75, 2);
            \draw[->, opacity=0.25] (-2, 2) -- (-1.75, 2);
            \draw[->, opacity=0.25] (-2, 1.8) -- (-1.75, 2);

            \draw[->, opacity=0.25] (-1.05, 1) -- (-0.05, 3);
            \draw[->, opacity=0.25] (-1.05, 1) -- (-0.05, 2);
            \draw[->, opacity=0.25] (-1.05, 1) -- (-0.05, 1);
            \draw[->, opacity=0.25] (-2, 1.4) -- (-1.75, 1);
            \draw[->, opacity=0.25] (-2, 1.2) -- (-1.75, 1);
            \draw[->, opacity=0.25] (-2, 1) -- (-1.75, 1);

            \draw[->] (0.65, 3) -- (1.65, 3);
            \draw[->] (0.65, 3) -- (1.65, 2);
            \draw[->] (0.65, 3) -- (1.65, 1);
            \draw[opacity=0.25] (2.35, 3) -- (2.6, 3);
            \draw[opacity=0.25] (2.35, 3) -- (2.6, 2.8);
            \draw[opacity=0.25] (2.35, 3) -- (2.6, 2.6);

            \draw[->] (0.65, 2) -- (1.65, 3);
            \draw[->] (0.65, 2) -- (1.65, 2);
            \draw[->] (0.65, 2) -- (1.65, 1);
            \draw[opacity=0.25] (2.35, 2) -- (2.6, 2.2);
            \draw[opacity=0.25] (2.35, 2) -- (2.6, 2);
            \draw[opacity=0.25] (2.35, 2) -- (2.6, 1.8);

            \draw[->] (0.65, 1) -- (1.65, 3);
            \draw[->] (0.65, 1) -- (1.65, 2);
            \draw[->] (0.65, 1) -- (1.65, 1);
            \draw[opacity=0.25] (2.35, 1) -- (2.6, 1.4);
            \draw[opacity=0.25] (2.35, 1) -- (2.6, 1.2);
            \draw[opacity=0.25] (2.35, 1) -- (2.6, 1);

            \draw[blue, opacity=0.8] (-0.15, 3.45) -- (0.75, 3.45) -- (0.75, 0.55) -- (-0.15, 0.55) -- cycle;
            \draw[->, blue, opacity=0.8] (0.3, 3.45) -- (0.3, 3.6) -- (-0.75, 3.6) -- (-0.75, 4.75) -- (-0.5, 4.75);
            \draw[->, blue, opacity=0.8] (1.5, 4.75) -- (2, 4.75) -- (2, 3.75);
            \draw[blue, opacity=0.8] (-0.5, 4) -- (-0.5, 5.5) -- (1.5, 5.5) -- (1.5, 4) -- cycle;

            \draw (-0.125, 4.25) -- (-0.125, 5) -- (0.875, 5) -- (0.875, 4.25) -- cycle;
            \draw (0.125, 4.5) -- (0.125, 5.25) -- (1.125, 5.25) -- (1.125, 4.5) -- cycle;
            \draw (-0.125, 4.25) -- (0.125, 4.5);
            \draw (-0.125, 5) -- (0.125, 5.25);
            \draw (0.875, 5) -- (1.125, 5.25);
            \draw (0.875, 4.25) -- (1.125, 4.5);
        \end{tikzpicture}
    \end{center}
    \caption{Bounding the solution space of variable vector $\mathbf{z}$ by the Viterbi update law and the pruning vector $\pmb{\eta}$ of Equation \eqref{eq:viterbi} defines a tropical polytope. The adaptive algorithm exploits the properties of the polytope to adapt the value of the leniency parameter $\theta$.}
    \label{fig:algo_poly}
\end{figure}
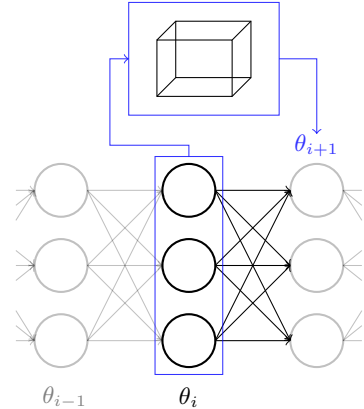

When the algorithm decides there exists a need for adaptation, the metric $\nu$ is examined and compared to its running history. This is done because, as already mentioned, $\nu$ is a measure of the volume of the solution space. Volume can convey quantitative information about the solution space. By maintaining that volume, we ensure that the solution space has some metric similarity throughout iterations. The normalized volume is useful since the values that the state vector admits change over time. As the total costs get larger (or the probabilities get smaller), then the leniency parameter $\theta$ has less of an effect. By trying to maintain the normalized volume, we ensure that the leniency parameter $\theta$ will adapt to accomodate larger weights in the later iterations of the Viterbi pruning. The algorithm tries to leverage these observations by keeping a history of the metric $\nu$. In cases were pruning is warranted (as indicated by the percentage change in $\varepsilon$), an effort is made to maintain the normalized volume metric $\nu$, in order to keep the size of the solution space similar. Figure \ref{fig:algo_poly} further illustrates this process.

Adapting the parameter based solely on the maintenance of the volume metric $\nu$ would have two undesirable effects:
\begin{itemize}
    \item Firstly, the parameter would be adapted at every iteration, in order to maintain the normalized volume. While this is not necessarily undesirable, it has certain implications. Examining only the volume metric we have a quantitative analysis of the solution space, but lack a qualitative analysis. This means that very few high probable paths are indistinguishable from a vast number of very low probability paths. We would like to be able to understand such a shift of the solution space, and volume-based metrics alone cannot offer such a luxury.
    \item Secondly, and most importantly, maintaining the starting volume is very strongly interweaved with the initial parameter $\theta_0$. Indeed, the volume is calculated based on the initial parameter $\theta_0$, and thus the whole premise of the algorithm would rely heavily on the value of that parameter.  While starting conditions still matter in the proposed algorithm, their effect is not as drastic as if we were to rely on the starting volume.
\end{itemize}
Thus, a combination of the metrics $\nu$ and $\varepsilon$ is considered in the adaptation of the leniency parameter $\theta$.

\section{Results}
\label{sec:res}
\begin{figure}[t]
    \begin{center}
        \includegraphics[width=0.5\textwidth]{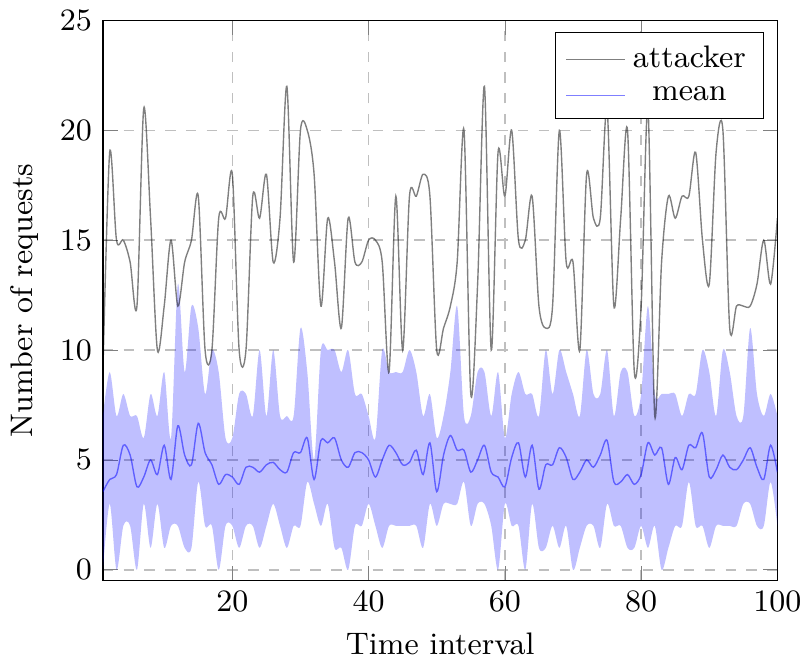}
    \end{center}
    \caption{Each user makes a number of requests to the network during different time intervals. The attacker makes a significantly higher number of requests at each interval compared to the average number of requests made by the benevolent users.}
    \label{fig:input_signal}
\end{figure}
We performed experiments to evaluate Algorithm \ref{algo:adapt} using a simulated attack on a network. In particular we assumed that a malevolent user has gained anauthorized access to the network and he wishes to disrupt its function by performing a large number of requests to the network, and thus hogging the available resources, hindering the service of benevolent users. We would like to utilise Algorithm \ref{algo:adapt} in order to solve the localisation problem and thus determine that malevolent user. We also assume that user exhibits some intelligence, by varrying the rate according to which he sends packets to the network in timed intervals. We also, in certain experiments, allow for the attacker to try and further spoof his identity by changing his position in the network, trying to further mask his identity. Thus, our adaptive algorithm will try to dynamically adapt the value of the pruning parameter $\theta$ across the iterations, in order to locate the attacker without wasting the system's resources (namely, without examining a large number of states). Figure \ref{fig:input_signal} represents the form of the input signals. There, the number of requests the attacker performs at each time interval is indicated by the black line. The average number of requests by all the other users in the network is indicated by the blue line. Also represented in the figure are the maximum and minimum numbers of requests the other users make to the network at each time interval. The number of requests for both the attacker and the benevolent users are modeled with Poisson distributions of varying parameters, as discussed in Section \ref{sec:back}.

In the modeling for the experiments, we assigned a lower Poisson parameter to the attacker. This choice was made because the original framework that was proposed for the metrics $\varepsilon$ and $\nu$ was modeled in min-plus algebra, and thus the Viterbi algorithm aims to find the sequence of the \emph{lowest cost}. In that vein, we decided to keep the modeling the same, to highlight the mathematical origin of the algorithm. At the same time, we wanted the experiment to maintain its immediately interpretable nature, meaning that the parameter of the Poisson distributions reflect the average number of requests by each user. Thus, we allowed a lower Poisson parameter to the attacker, and perform the min-plus calculations.

\begin{figure}[t]
    \begin{center}
        \includegraphics[width=0.5\textwidth]{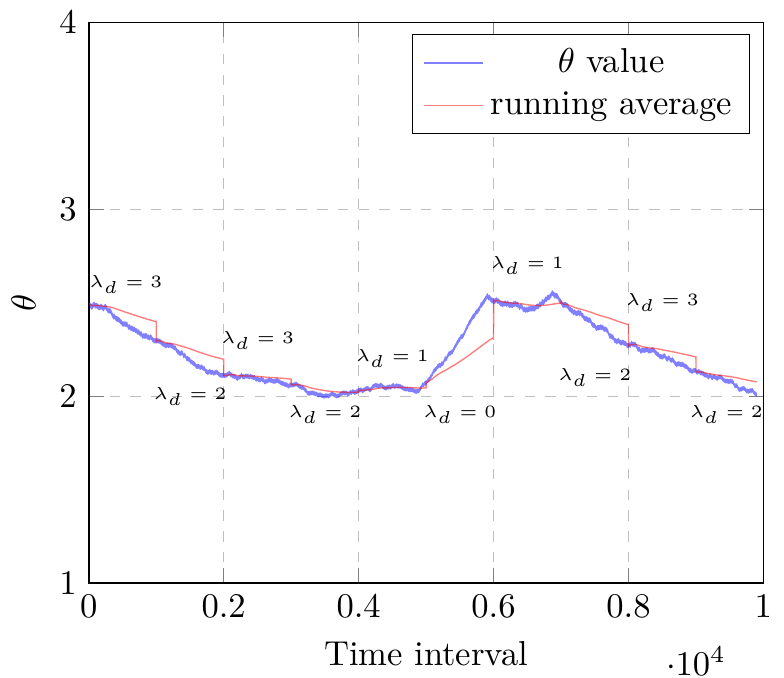}
    \end{center}
    \caption{Algorithm \ref{algo:adapt} adapts the value of the leniency parameter $\theta$ at each time frame according to the metrics $\nu$ and $\varepsilon$ (Equations \eqref{eq:nu} and \eqref{eq:varepsilon}). The attacker changes the rate of his requests every $1000$ iterations.}
    \label{fig:theta_adapt}
\end{figure}

Figure \ref{fig:theta_adapt} presents how the algorithm adapts the leniency parameter $\theta$ for an instance of the experiment. The particular values of that experiment were $\alpha = 0.25$, $\beta = 0.0005$, and $\theta_0 = 2.5$. The value $\beta$ is similar to the learning rate of optimisation algorithms, and thus needs to be sufficiently small, in order to regulate the weight each iteration has.

\section{Conclusion}
\label{sec:conc}
In this work we proposed a variation of the Viterbi pruning as the solution for an attacker localisation problem. In particular, we proposed an adaptive pruning algorithm inspired by the geometrical aspect of the tropical analysis of the Viterbi pruning. By analysing the tropical geometry of the traditional pruning algorithm, we incorporate metrics into the proposed adaptive algorithm in order to evaluate the need for adaptation. In the case the algorithm deemed that the current time interval's metrics vary sufficiently from the previous history, then an adaptation is made to the effect of maintaining the previous levels of the enclosed volume $\nu$. We experimented with various values for the parameters of the algorithm and presented numerical results of the application of the proposed algorithm in the task of locating a simulated attacker on a network.

\newpage


\printbibliography

\end{document}